\documentclass[fleqn,12pt,twoside]{article}
\usepackage{espcrc1}

\usepackage{epsfig}
\usepackage[figuresright]{rotating}

\newcommand{\AmS}{{\protect\the\textfont2
  A\kern-.1667em\lower.5ex\hbox{M}\kern-.125emS}}

\newcommand{\beq}{\begin{equation}}
\newcommand{\eeq}{\end{equation}}
\newcommand{\bea}{\begin{eqnarray}}
\newcommand{\eea}{\end{eqnarray}}
\newcommand{\ben}{\begin{eqnarray*}}
\newcommand{\een}{\end{eqnarray*}}

\newcommand{\simle}{\hspace*{0.2em}\raisebox{0.5ex}{$<$}
     \hspace{-0.8em}\raisebox{-0.3em}{$\sim$}\hspace*{0.2em}}

\newcommand{\boldtau}{\mbox{\boldmath $\tau$}}

\newcommand{\bolds}{\mbox{\boldmath $s$}}

\newcommand{\nab}{\overrightarrow{\nabla}}

\newcommand{\Mlo}{M_{\rm lo}}
\newcommand{\Mhi}{M_{\rm hi}}

\title{Effective Field Theories of Light Nuclei}

\author{U. van Kolck\address[UA]{Department of Physics, 
        University of Arizona\\ 
        Tucson, AZ 85721, USA}%
        \thanks{Supported in part by the US Department of Energy and
                the Alfred P. Sloan Foundation.}}
       
\begin{document}

\maketitle

\begin{abstract}
Effective field theories have been developed for the description
of light, shallow nuclei.
I review results for two- and three-nucleon systems, and discuss their
extension to halo nuclei.
\end{abstract}

\section{EFFECTIVE FIELD THEORIES}

I will remember G\"oteborg as a clean and ordered town.
INPC 2004 was certainly well organized. My talk, too,  was about organization.

Nuclear structure involves energies that are
much smaller than the typical QCD mass scale, $M_{QCD}\sim 1$ GeV.
This is a common situation in physics: 
an ``underlying'' theory is valid at a mass scale
$M_{\rm hi}$, but we want to study processes at
momenta $Q$ of the order of a lower scale 
$M_{\rm lo}\ll M_{\rm hi}$.
Typically, there is ``more'' at lower energies.
How to organize the complexity brought in 
by the ``effective'' interactions that will ensure
that low-energy observables are described correctly?

Effective Field Theory  (EFT) is a framework to
construct these interactions systematically,
at the same time maintaining desirable general principles
such as causality and cluster decomposition.
Here I discuss the application of EFT to a class
of nuclear systems: those that exhibit poles
in the complex momentum plane at a scale much smaller than 
the pion mass, that is, $M_{\rm hi}\simle m_\pi$.
They include two- and three-nucleon systems, and 
other halo nuclei.

EFT starts with the observation that the effective interactions
consist of the sum of {\it all} possible interaction terms
in a Lagrangian that involves only the 
fields representing
low-energy degrees of freedom.
Because of the uncertainty principle,
each of these interaction terms can be taken as a local combination
of derivatives of the fields.
If the ``integrating out'' of 
the high-energy degrees of freedom is done appropriately,
the effective Lagrangian will have the same symmetries as the 
underlying theory.
The details of the underlying dynamics, on the other hand, 
are contained in the interaction
strengths. The latter depend also on the details
of how the low- and high-energy degrees of freedom are separated.
This separation requires the introduction of a cutoff parameter $\Lambda$
with dimensions of energy.
Both the interaction
strengths and  the quantum effects represented by loops
depend on $\Lambda$.
However, the cutoff procedure is arbitrary, so by construction observables
are independent of $\Lambda$ (``renormalization-group invariance'').
The $T$ matrix for any low-energy process acquires the schematic form
\begin{equation}
T (Q\sim \Mlo)= {\cal N} \sum_{\nu=\nu_{\rm min}}^\infty c_\nu (\Mhi, \Lambda) 
               \left(\frac{Q}{\Mhi}\right)^\nu
               {\cal F}_\nu \left(\frac{Q}{\Mlo}; \frac{\Lambda}{\Mlo}\right),
\label{T}
\end{equation}
where ${\cal N}$ is a common normalization factor, 
$\nu$ is a counting index starting at some value $\nu_{\rm min}$,
the $c_\nu$'s are parameters, 
and the ${\cal F}_\nu$'s are calculable functions.
We must have
\begin{equation}
\frac{\partial T(Q\sim \Mlo)}{\partial \Lambda}=0.
\label{delT}
\end{equation}

In order to maintain predictive power in the effective theory
it is necessary to truncate the sum in Eq. (\ref{T}) in such a way
that the resulting cutoff dependence can be decreased
systematically with increasing order.
We call such ordering ``power counting''.
There are essentially two ways of doing this.
One is to carry out the integration of high-energy degrees of freedom
explicitly
---as it is done in going from QCD to the effective
hadronic theory through lattice simulations---
and infer the power counting from the sizes of the calculated
terms.
Another, which we use when we do not know or cannot solve the
underlying theory, is to guess the sizes of the effective
interactions by assuming that the
renormalized interactions are natural,
that is, are in order of magnitude given by
$\Mhi$ to a power determined by dimensional analysis.
This guess is confirmed {\it a posteriori}, by checking
renormalization-group invariance and convergence of the truncation
after the data is fitted order by order.
In some cases, including the ones considered here,
there exists a fine-tuning requires a bit more thought.

For the last ten years or so we have been developing 
EFTs for systems of few nucleons
\cite{ARNPSreview}.
The goal is to understand traditional nuclear physics
from a QCD standpoint.
Much of this work has been devoted to the EFT where 
$M_{\rm hi}\sim M_{QCD}$ and $M_{\rm lo}\sim m_\pi$ 
\cite{ARNPSreview,ulfstalk}.
In this EFT, pions are explicit degrees of freedom,
and (approximate) chiral symmetry plays a crucial role.
While in the sector of $A=0, 1$ nucleons this ``pionful'' EFT reduces
to well-understood chiral perturbation theory, 
in the $A\ge 2$ sector power counting
is more subtle \cite{BBSvk}.
Nevertheless, a reasonably successful potential has been constructed
\cite{ARNPSreview,ulfstalk}.

Now, in many situations ---in particular, many astrophysical applications---
we are interested in reactions at momenta $Q\ll m_\pi$.
Moreover, as we are going to see, 
there is interesting nuclear physics in this regime.

\section{TWO AND THREE-NUCLEON SYSTEMS}

The typical momentum of nucleons in the deuteron
is $\aleph_1\sim \sqrt{m_N B_d}\simeq 45$ MeV, which means that 
the deuteron is an object about three times larger than most of the pion cloud
around each nucleon.
For the $s_0$ virtual bound state, the corresponding scale is
even smaller, $\aleph_0\sim \sqrt{m_N B_d'}\simeq 8$ MeV.
For these states the pionful EFT is an overkill.
One can instead consider a much simpler EFT where the meson cloud 
is represented by a multipole expansion: the Lagrangian contains
only nucleon fields with contact interactions.
 
This ``pionless'' EFT, for which
$M_{\rm hi}\sim m_\pi$ and $M_{\rm lo}\sim \aleph$
(with $\aleph$ some average of the $\aleph_i$),
is now pretty well understood,
despite the fact that one needs to account for fine-tuning
through the anomalously-small scale $\aleph$.
(It seems accidental that $\aleph$ is smaller than a pion
scale such as $4\pi f_\pi^2/m_N$, which arises naturally in the
pionful EFT \cite{ARNPSreview}.)
One way to do this is to introduce, in addition to the nucleon spinor/isospinor
field $N$ of mass $m_N$,
two auxiliary ---``dimeron''--- fields, a 
scalar/isovector $\bolds_0$ and 
a vector/isoscalar $\vec{s}_1$ 
with masses $\Delta_i$ \cite{Kap97}.
(The final results for observables are, of course,
independent of the choice of fields.)
The most general 
parity- and time-reversal-invariant Lagrangian is 
\cite{pionlessPC,gautam,3bosons,pionlessNd,pionlessT}
\begin{eqnarray}
{\cal L}&=&
N^\dagger \left(i\partial_0+ \frac{\vec{\nab}^{\,2}}{2m_{N}}\right)N
+\Delta_{0} \, \bolds_0^\dagger \cdot \bolds_0
+\Delta_{1} \, \vec{s}_1^{\, \dagger}\cdot \vec{s}_1
\nonumber\\
&& -\frac{g_{0}}{2}\left[ \bolds_0^\dagger \cdot N^T \sigma_2 \boldtau\tau_2 N
+{\rm H.c.}\right]
-\frac{g_{1}}{2}\left[ \vec{s}_1^{\, \dagger} \cdot 
                              N^T \tau_2\vec{\sigma} \sigma_2 N
+{\rm H.c.}\right]
\nonumber\\
&& 
-h \left\{g_{0}^2 N^\dagger (\bolds_0\cdot\boldtau)^\dagger 
                    (\bolds_0\cdot\boldtau) N 
+g_{1}^2 N^\dagger  (\vec{s}_1\cdot\vec{\sigma})^\dagger 
                    (\vec{s}_1\cdot\vec{\sigma}) N 
\right.\nonumber\\
&& \left.\qquad
+\frac{g_{0}g_{1}}{3}\left[N^\dagger
                    (\vec{s}_1\cdot\vec{\sigma})^\dagger
                    (\bolds_0\cdot\boldtau) N +{\rm H.c.}\right]\right\}
+\ldots,
\label{lag1}
\end{eqnarray}
where the $g_i$ and $h$ are coupling constants
to be determined.
In addition to the kinetic terms for the various fields,
the Lagrangian contains all interactions between nucleon
and auxiliary fields. 
(Integrating over the auxiliary fields
in the  path integral results in a completely equivalent form
of the EFT, without auxiliary fields and with purely-contact interactions.)
Only some of the most important terms are shown explicitly here:
those that contribute to the $s$ waves in the two- and three-nucleon systems.
The ``$\dots$'' include terms with more derivatives 
and contributions to other waves.

\subsection{The two-nucleon system}

\begin{figure}[tb]
\centerline{\psfig{file=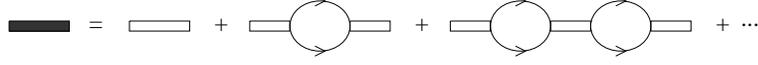,width=10cm}}
\vspace*{8pt}
\caption{The full dimeron propagator (thick shaded line) is
obtained by dressing the bare dimeron propagator (double solid
line) with particle bubbles (solid lines) to all orders.}
\label{fig:auxprop}
\end{figure}

In the two-nucleon ($NN$) system, 
the full $T$ matrix can be obtained by adding nucleon legs to the
full dimeron propagators.
The latter consist of insertions of particle bubbles generated by the
two-particle/dimeron interaction ---see Fig.\ref{fig:auxprop}---
as well as insertions stemming from terms with more derivatives. 
Since a two-nucleon bubble is ${\cal O}(m_N Q/4\pi)$,
if $\Delta_i= {\cal O}(\aleph_i)$ and
$g_i^2/4\pi={\cal O}(1/m_N)$, then the bubbles have to be resummed in the 
$s$ waves \cite{pionlessPC}.
That is, 
with the dimeron masses fine-tuned, the leading interactions 
have to be summed to all orders.
The $T$ matrix develops poles at $Q\sim \pm i\aleph_i$,
which correspond to 
the observed shallow bound states.
Higher-derivative interactions are smaller by powers of
$Q/M_{\rm hi}$.

\begin{figure}[tb]
\begin{minipage}[t]{80mm}
\centerline{\psfig{file=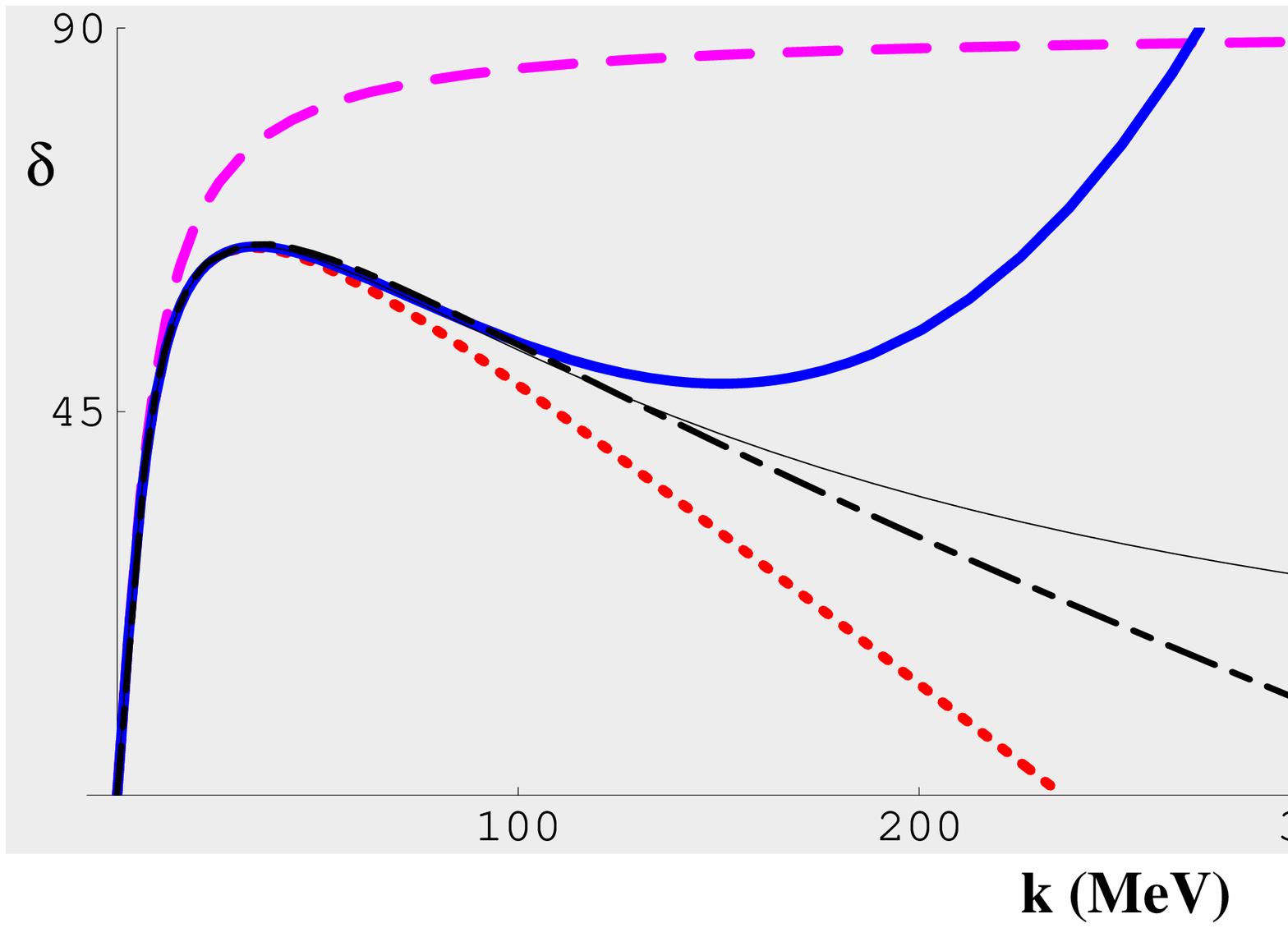,width=8cm,clip=true}}
\end{minipage}
\hspace{\fill}
\begin{minipage}[t]{75mm}
\centerline{\psfig{file=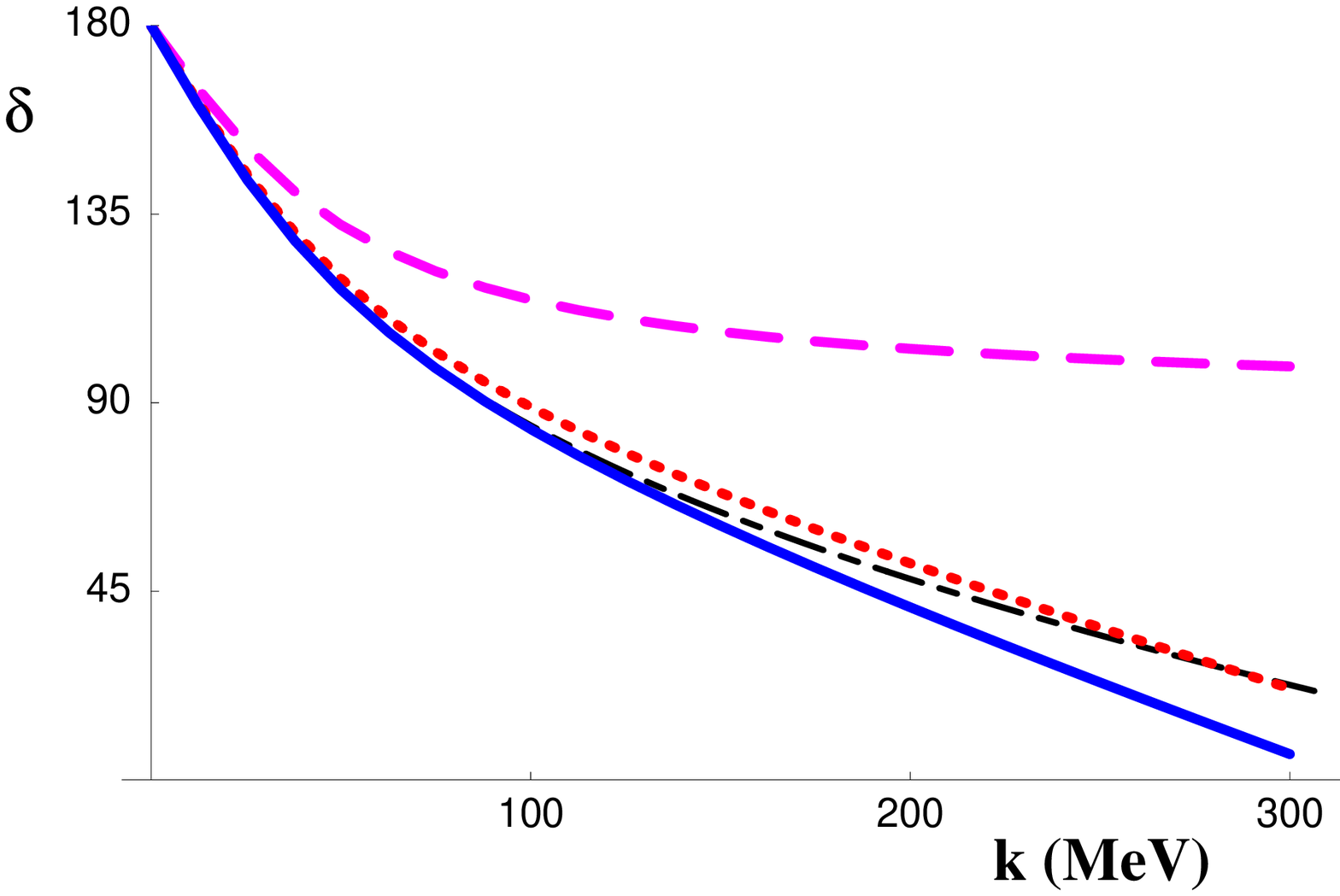,width=8cm,clip=true}}
\end{minipage}
\caption{The $s_0$ (left) and $s_1$ (right) $NN$ phase shifts
(in degrees)
as functions of the center-of-mass momentum (in MeV).
The dash-dotted lines are 
the Nijmegen phase-shift analysis \protect\cite{Nijm}.
Left:
the dashed, dotted, and thick solid lines show the EFT results 
at LO, NNLO, and NNNNLO, respectively, while the
thin solid line shows the effective-range result. 
Right: 
the dashed, dotted, and thick solid lines show the EFT results 
at LO, NLO, and NNLO, respectively. 
From Refs.~\protect\cite{ARNPSreview,gautam}, courtesy of M. Savage.}
\label{fig:NNphases}
\end{figure}

The $NN$ $T$ matrix 
has the form (\ref{T}) with 
${\cal N}=4\pi/\mu \Mhi$ and 
$\nu= \sum_i V_i d_i -p +L$,
where in a diagram 
$V_i$ is the number of vertices with  $d_i$ derivatives,
$p$ is the number of $s_0$ or $s_1$ propagators, 
and $L$ is the number of loops.
One can show  \cite{pionlessPC} that it 
is equivalent order by order to
that of the effective-range expansion.
$T$ involves in leading order (LO) only the scattering lengths
$|a_{i}|\sim 1/\aleph_i$;
at next-to-leading order (NLO) and next-to-next-to-leading order (NNLO), 
also the effective ranges $|r_{i}| \sim 1/m_\pi$;
and so on.
The resulting phase shifts  \cite{gautam,ARNPSreview}
converge to empirical values
for $Q\simle m_\pi$,
examples being shown in Fig. \ref{fig:NNphases}.
The deuteron binding energy is found to be $B_d=1.91$ MeV in NLO,
to be compared with the experimental value of 2.22 MeV.

In addition, many low-energy reactions involving the deuteron have been 
studied with this EFT ---see Ref. \cite{ARNPSreview} for a review.

\subsection{The three-nucleon system}

\begin{figure}[tb]
\begin{minipage}[t]{75mm}
\centerline{\psfig{file=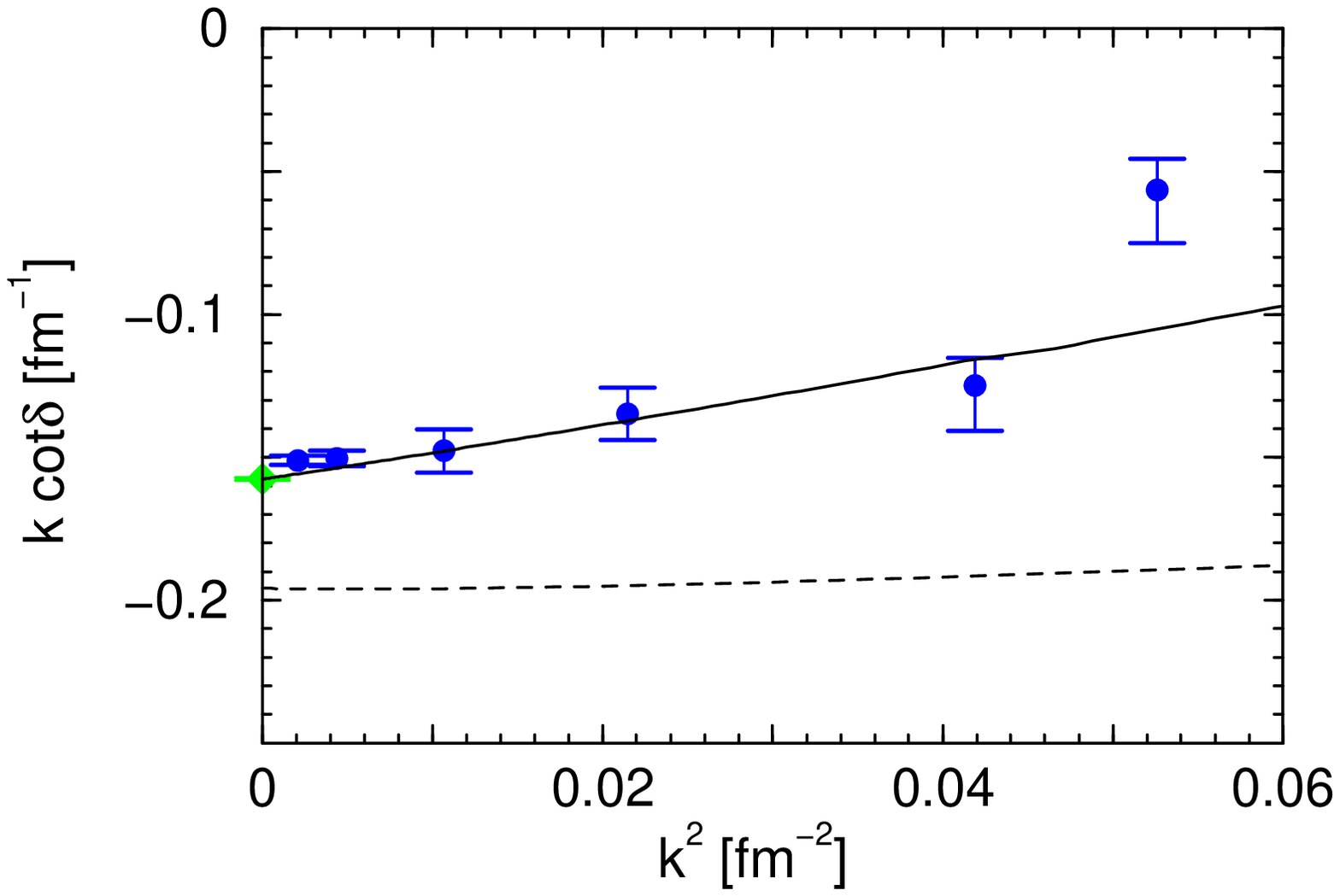,width=8cm,clip=true}}
\end{minipage}
\hspace{\fill}
\begin{minipage}[t]{80mm}
\centerline{\psfig{file=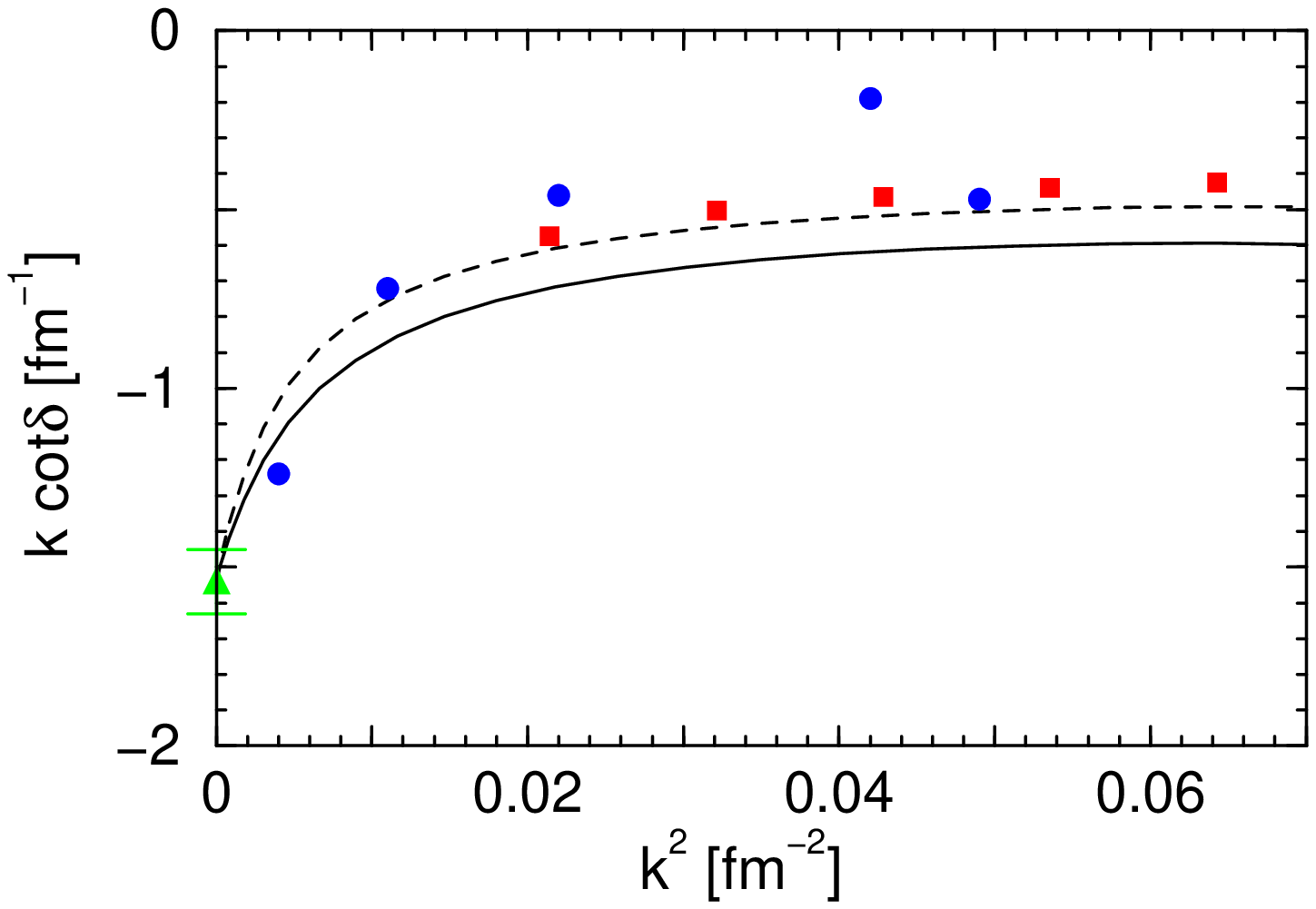,width=7.75cm,clip=true}}
\end{minipage}
\caption{The $s_{3/2}$ (left) and $s_{1/2}$ (right) $Nd$ $K^{-1}$ matrix 
as function of the square of the center-of-mass momentum (in fm$^{-2}$).
The points at threshold are from a cold-neutron measurement \protect\cite{CN}
and the dots from a phase-shift analysis \protect\cite{vOS}.
Left: 
dashed and solid lines show the EFT results at LO and NNLO, respectively. 
Right:
solid and dashed lines show the EFT results at LO and NLO, respectively,
while the squares are the result of a phenomenological potential model 
\protect\cite{kievsky}. 
From Refs.~\protect\cite{pionlessNd,pionlessT},
courtesy of H. Hammer.}
\label{fig:Ndphases}
\end{figure}

The three-nucleon ($3N$) system is more interesting.
In all but the $s_{1/2}$ wave, $3N$ forces appear only 
at high orders, and very precise results for 
nucleon-deuteron ($Nd$) scattering follow with
parameters fully determined from $NN$ scattering \cite{pionlessNd}.
The $s_{3/2}$ phase shift is given as an example
in Fig. \ref{fig:Ndphases}:
excellent agreement with data is achieved already at NNLO.
In particular, the scattering length is postdicted
as $a_{3/2}=6.33\pm 0.10$ fm, to be compared to 
the experimental value, $6.35\pm 0.02$ fm.
One can, thus, do QED-quality nuclear physics with EFT.

In the $s_{1/2}$ wave, renormalization-group 
invariance can only be achieved if the 
$3N$ interactions 
are also enhanced by 
two powers of $\aleph^{-1}$ \cite{3bosons,pionlessT}.
In this channel, a single non-derivative $3N$ interaction
appears in LO, and higher-derivative interactions 
are smaller by powers of
$Q/M_{\rm hi}$. 
To NLO there is only one parameter
not fixed by $NN$ observables ---$h$ in Eq. (\ref{lag1})).
It can be fixed by, say, the  $s_{1/2}$ $Nd$ scattering length,
and as a function of the cutoff it displays an unusual,
limit-cycle behavior.
The resulting energy dependence of 
$s_{1/2}$ $Nd$ scattering comes
out very well \cite{pionlessT},
see Fig. \ref{fig:Ndphases}. 
Likewise, the triton binding energy is found to be $B_t=8.31$ MeV in NLO,
to be compared with the experimental value of 8.48 MeV.

This EFT simplifies the treatment of light nuclei,
but the application to larger nuclei still faces computational
challenges. 
(For the first attack on the four-body system, 
see Refs. \cite{4bosons,four}.)
One would like to devise further simplifications
in order to extend EFTs to larger nuclei.
As a first step, we can specialize
to very low energies where clusters of nucleons 
behave coherently. Even though many interesting issues of 
nuclear structure are missed, we can at least describe 
anomalously-shallow (``halo'') nuclei and some reactions of 
astrophysical interest.

\section{HALOS}

I define a halo system as one that contains
two momentum scales:
\begin{itemize}
\item $M_{\rm hi}\sim \sqrt{m_N E_{\rm core}}$, associated with the 
excitation energy $E_{\rm core}$ of a tight cluster of nucleons (``core'');
\item $M_{\rm lo}\sim \sqrt{m_N E_{\rm halo}}$, 
associated with the 
energy $E_{\rm halo}$ for the attachment or removal of 
one or more (``halo'') nucleons.
\end{itemize}
These systems exhibit shallow $S$-matrix poles,
either on the imaginary axis (bound states)
or elsewhere in the complex momentum plane (resonances).
With this definition, the deuteron and the triton
are two- and three-body halo systems, respectively.
In these cases the core is a single nucleon,
$E_{\rm core}\sim m_\pi^2/m_N$, while
$E_{\rm halo}\sim B_d$ ($B_t$) for the deuteron (triton).
The next-simplest examples involve a $^4$He core,
for which $E_{\rm core}\simeq 20$ MeV.
In contrast, the removal energy for two neutrons from $^6$He
is $E_{\rm halo}\simeq 1$ MeV \cite{tunl},
making this a three-body halo nucleus.
It is interesting that $^5$He is not bound.
However,
the total cross section
for neutron-alpha ($n\alpha$) scattering
has a prominent bump at $E_{\rm halo}\sim 1$ MeV,
usually interpreted as a shallow $p_{3/2}$ resonance \cite{tunl}.
In addition, 
reactions involving more complex nuclei are frequently characterized
by shallow resonances that are narrow,
corresponding to poles near the real 
momentum axis. 

It is natural to generalize the EFT to describe shallow 
two-body resonances \cite{halos,resonances},
as a step before tackling three-body halo nuclei.
Here, for concreteness, 
I consider $n\alpha$ scattering,
which will fix the $N\alpha$ effective interactions,
necessary for a future study of $^6$He.

Now, in addition to a nucleon field,
I need to consider also a scalar/isoscalar $\phi$ field to represent
the $^4$He core 
of mass
$m_\alpha$.
I also introduce isospinor dimeron fields 
$s$, $d$, $t$, {\it etc.} 
with masses $\Delta_{0+}$, $\Delta_{1-}$, $\Delta_{1+}$, {\it etc.}, 
which
can be thought of as bare fields for the various
$N\alpha$ channels:
$s_{1/2}$, $p_{1/2}$, $p_{3/2}$, {\it etc.}, which I denote
$0+$, $1-$, $1+$, {\it etc.}
The most general
parity- and time-reversal-invariant Lagrangian is \cite{halos,resonances}
\begin{eqnarray}
{\cal L}&=&\phi^\dagger \left(i\partial_0 + 
\frac{\nab^{\,2}}{2m_{\alpha}}\right)\phi 
+N^\dagger \left(i\partial_0 + \frac{\nab^{\,2}}{2m_{N}}\right)N
+  t^\dagger \left(i\partial_0+\frac{\nab^{\,2}}{2(m_\alpha+m_N)}
+\Delta_{1+} \right) t \nonumber\\
&& +\frac{g_{1+}}{2}\bigg\{ t^\dagger \vec{S}^{\, \dagger}\cdot
\bigg[N \nab \phi-(\nab N)\phi\bigg]+ {\rm H.c.} 
\bigg\}\nonumber\\
&& + \Delta_{0+} \, s^\dagger s +g_{0+} \bigg[s^\dagger N \phi 
+\phi^\dagger N^\dagger s \bigg]
+  g_{1+}'  t^\dagger\left(i\partial_0+\frac{\nab^{\,2}}{2(m_\alpha+m_N)}
\right)^2 t
+\ldots,
\label{lag2}
\end{eqnarray}
in a notation similar to the one used in Eq. (\ref{lag1}),
where additionally the $S_i$'s 
are standard $2\times 4$ spin-transition
matrices connecting states with total angular momentum
$j=1/2$ and $j=3/2$. 
Again, of all possible interactions among nucleon, alpha
and auxiliary fields,
only some of the most important terms are shown explicitly here:
those that contribute to the $p_{3/2}$ and $s_{1/2}$ partial waves.
The ``$\dots$'' include terms with more derivatives 
and contributions to other waves.

The $N\alpha$ $T$ matrix 
can be obtained from the full dimeron
propagators by attaching external nucleon and alpha legs. 
The bubbles in the dressing of dimeron
propagators ---see Fig.\ref{fig:auxprop} again---
now represent the propagation of a nucleon
and an alpha particle.

\subsection{Low energies}

A bare dimeron propagator
can generate two shallow real poles provided its $\Delta$ is 
very small:
I take
$\Delta_{1+} \sim M_{\rm lo}^2/\mu$ 
(where $\mu$ is the $N\alpha$ reduced mass).
The bubbles introduce unitarity corrections, which
can dislocate the poles to the lower half-plane.
The resonance
will be narrow if the EFT is perturbative
in the coupling $g_{1+}$.
This will be so if 
$ g_{1+}^2/4\pi\sim 1/M_{\rm hi}\mu^2$, in which case a loop is suppressed
by $M_{\rm lo}/M_{\rm hi}$. 
Higher-derivative terms 
will also be 
perturbative if their strengths scale with $M_{\rm hi}$ 
according to their mass dimensions.
Likewise, parameters in waves without shallow resonances
will be given solely in terms of $M_{\rm hi}$, 
{\it e.g.}  $\Delta_{0+} \sim M_{\rm hi}$.
This EFT then describes a shallow, narrow resonance
with a single fine-tuned parameter $\Delta_{1+}$.

With this scaling, the $N\alpha$ $T$ matrix has
the form (\ref{T}) with 
${\cal N}=4\pi/\mu \Mhi$ and 
$\nu= \sum_i V_i d_i -2p +L$,
where 
now $p$ is the number of $1+$ propagators.
As a consequence,
$T$ involves in LO the scattering ``lengths''
$|a_{0+}|\sim 1/\Mhi$ and $|a_{1+}|\sim 1/\Mhi\Mlo^2$,
and the effective ``range'' $|r_{1+}| \sim \Mhi$ only;
at NLO, the unitarity corrections
in the same $0+$ and $1+$ waves;
at NNLO, 
$|r_{0+}| \sim 1/\Mhi$, 
the shape parameter $|{\cal P}_{1+}| \sim \Mhi^3$,
and 
$|a_{1-}|\sim 1/\Mhi^3$; and so on.

\begin{figure}[tb]
\begin{minipage}[t]{75mm}
\centerline{\psfig{file=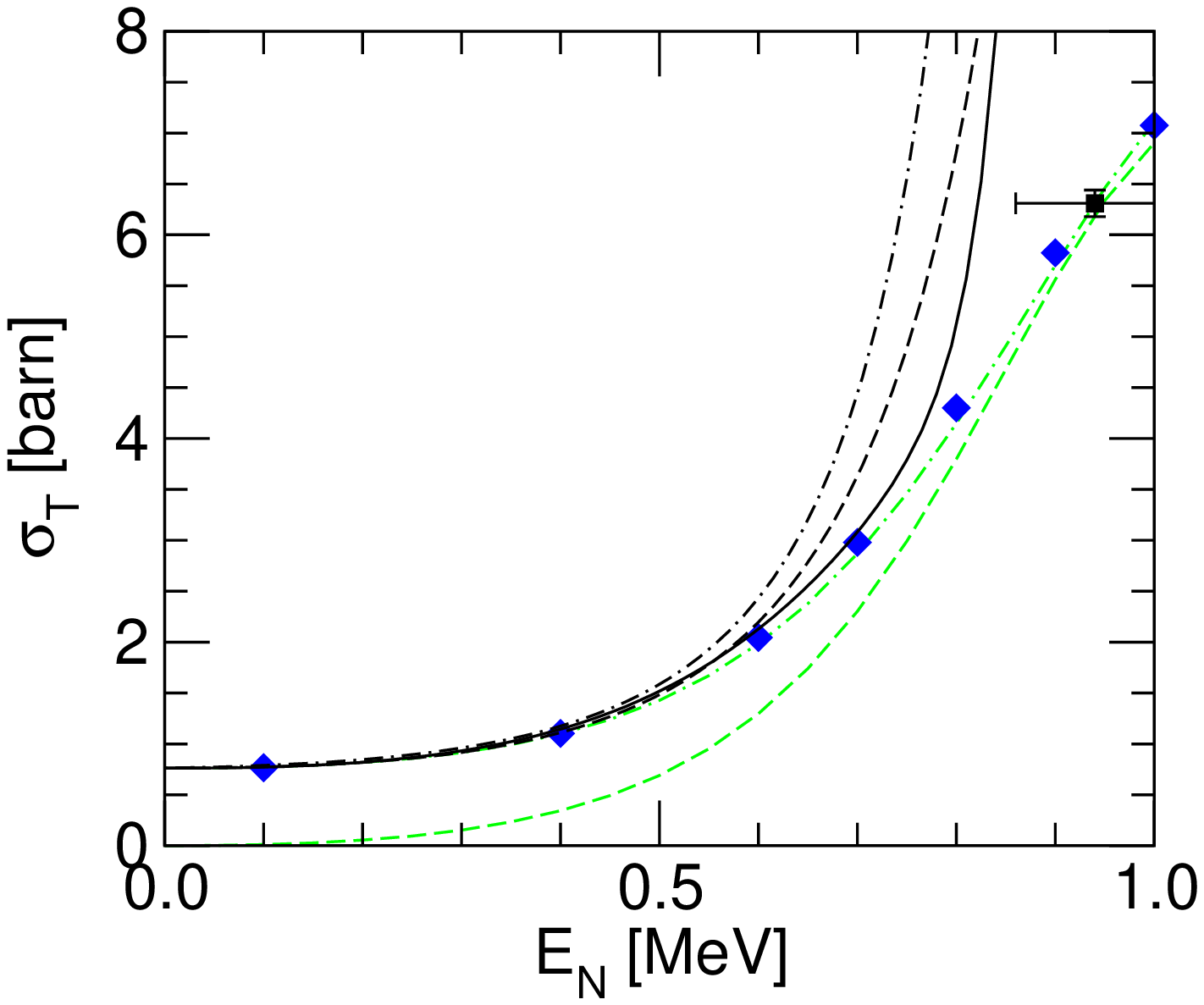,width=7.5cm,clip=true}}
\end{minipage}
\hspace{\fill}
\begin{minipage}[t]{80mm}
\centerline{\psfig{file=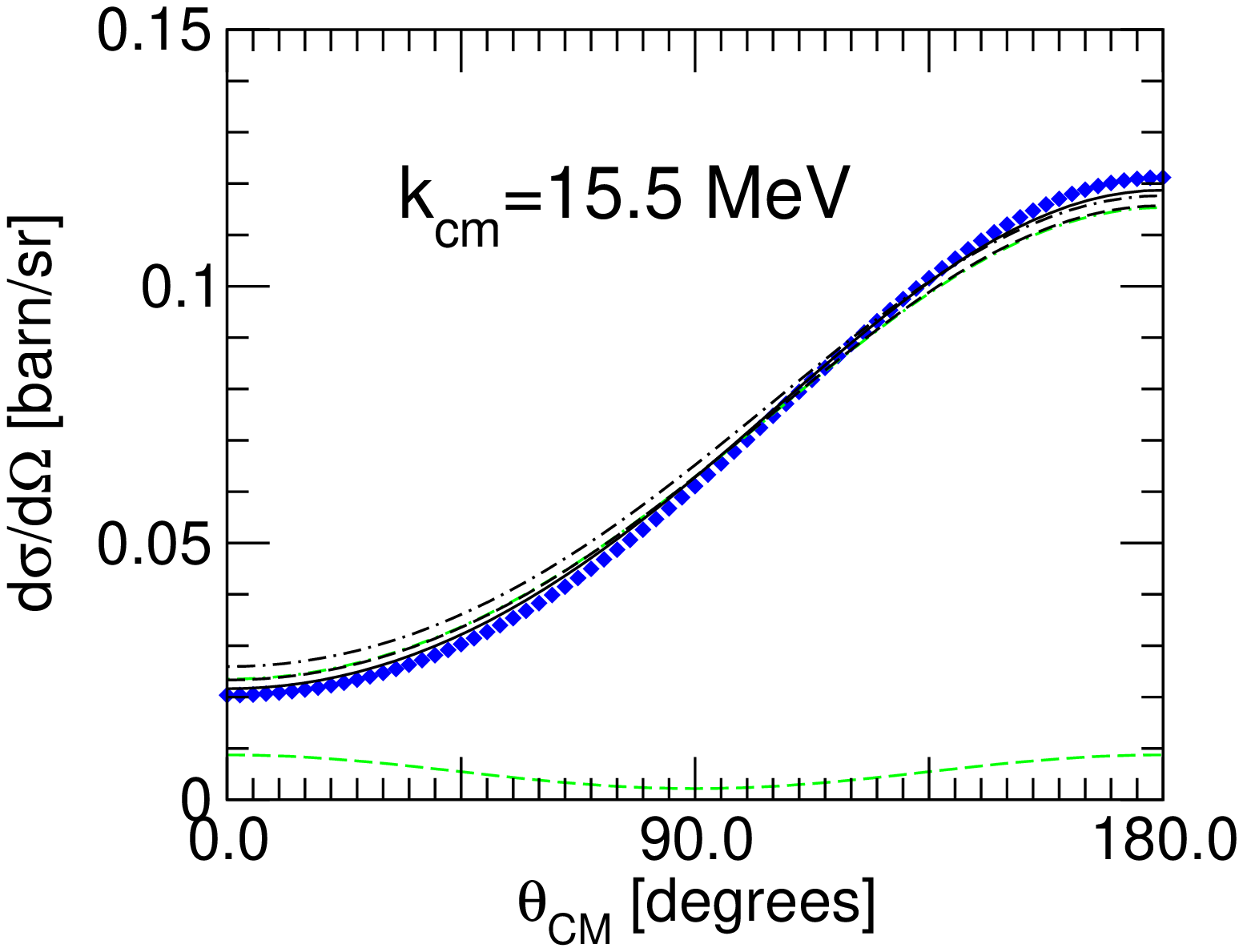,width=8cm,clip=true}}
\end{minipage}
\caption{Cross sections for $n\alpha$ scattering below the $p_{3/2}$
 resonance.
 Left: total cross section (in barns) as a function of the neutron kinetic 
 energy  (in MeV) in the $\alpha$ rest frame. 
 Right: differential cross section (in barns/sr) 
 as a function of the center-of-mass scattering angle $\theta_{cm}$ 
 (in degrees) at a center-of-mass momentum
 of $k_{cm}=15.5$ MeV. 
 The diamonds are evaluated data \protect\cite{BNL}, 
 and the black squares are experimental 
 data \protect\cite{data}.
 The dashed, dash-dotted, and solid black lines show the EFT result
 without resummation at LO, NLO, and NNLO, respectively. The 
 gray dashed and dash-dotted lines show the EFT result with  
 resummation at LO and NLO, respectively.
From Ref.~\protect\cite{resonances}.}
\label{fig:sigs1}
\end{figure}

We fit the EFT parameters to an $n\alpha$ phase-shift analysis \cite{ALR73},
and find $\Mhi\sim 100$ MeV and $\Mlo\sim 30$ MeV.
The results \cite{resonances}
for the total and differential $n\alpha$ cross sections
are compared with data in Fig. \ref{fig:sigs1}.
The data are reproduced up to neutron
energies of about $E_N\approx 0.5$ MeV in LO and
$0.8$ MeV in NNLO.
(Interestingly, the NLO result worsens the 
description of the data.) 
The expansion fails in the immediate neighborhood of the resonance.

\subsection{Around the resonance}

\begin{figure}[tb]
\begin{minipage}[t]{75mm}
\centerline{\psfig{file=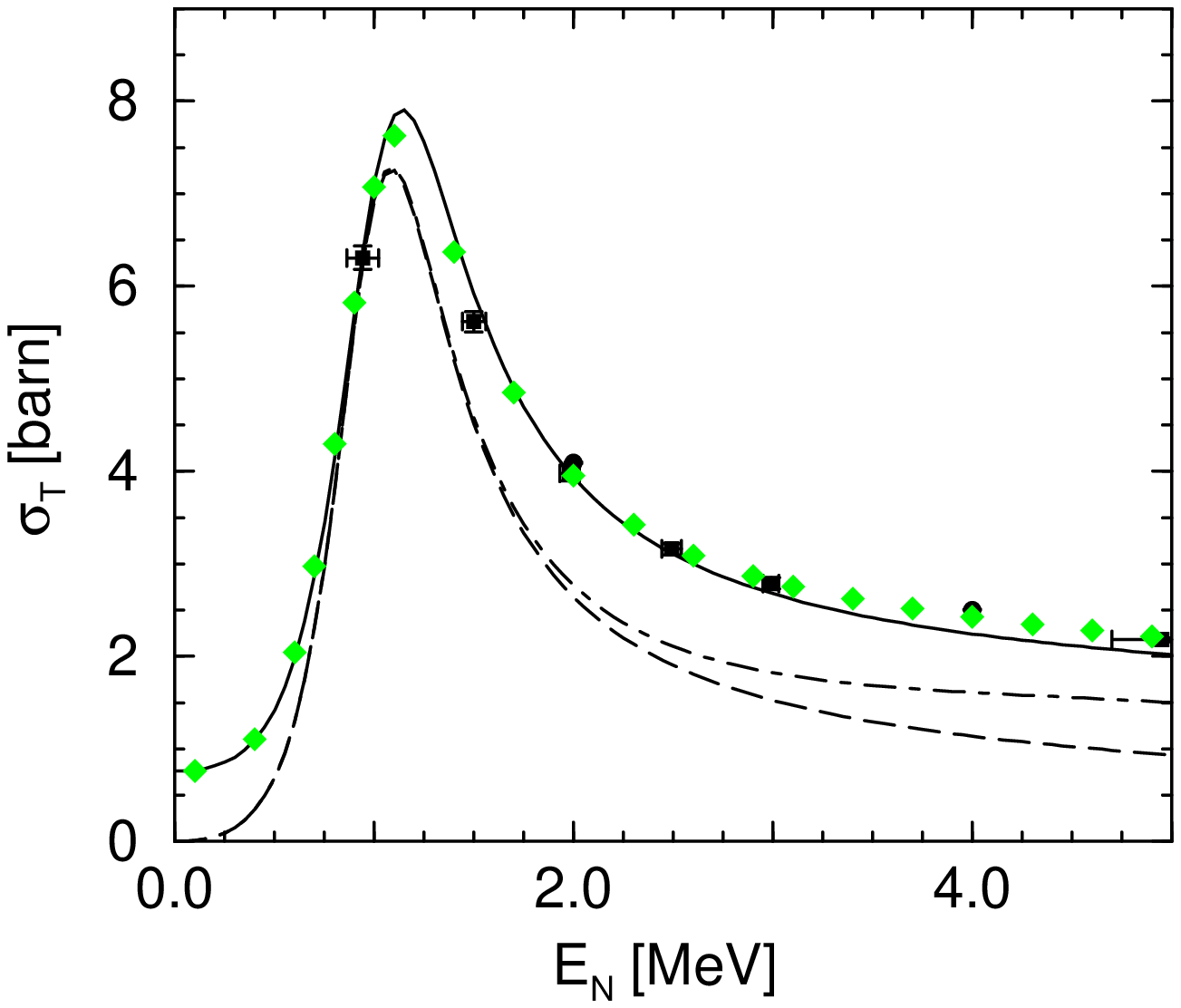,width=7.25cm,clip=true}}
\end{minipage}
\hspace{\fill}
\begin{minipage}[t]{80mm}
\centerline{\psfig{file=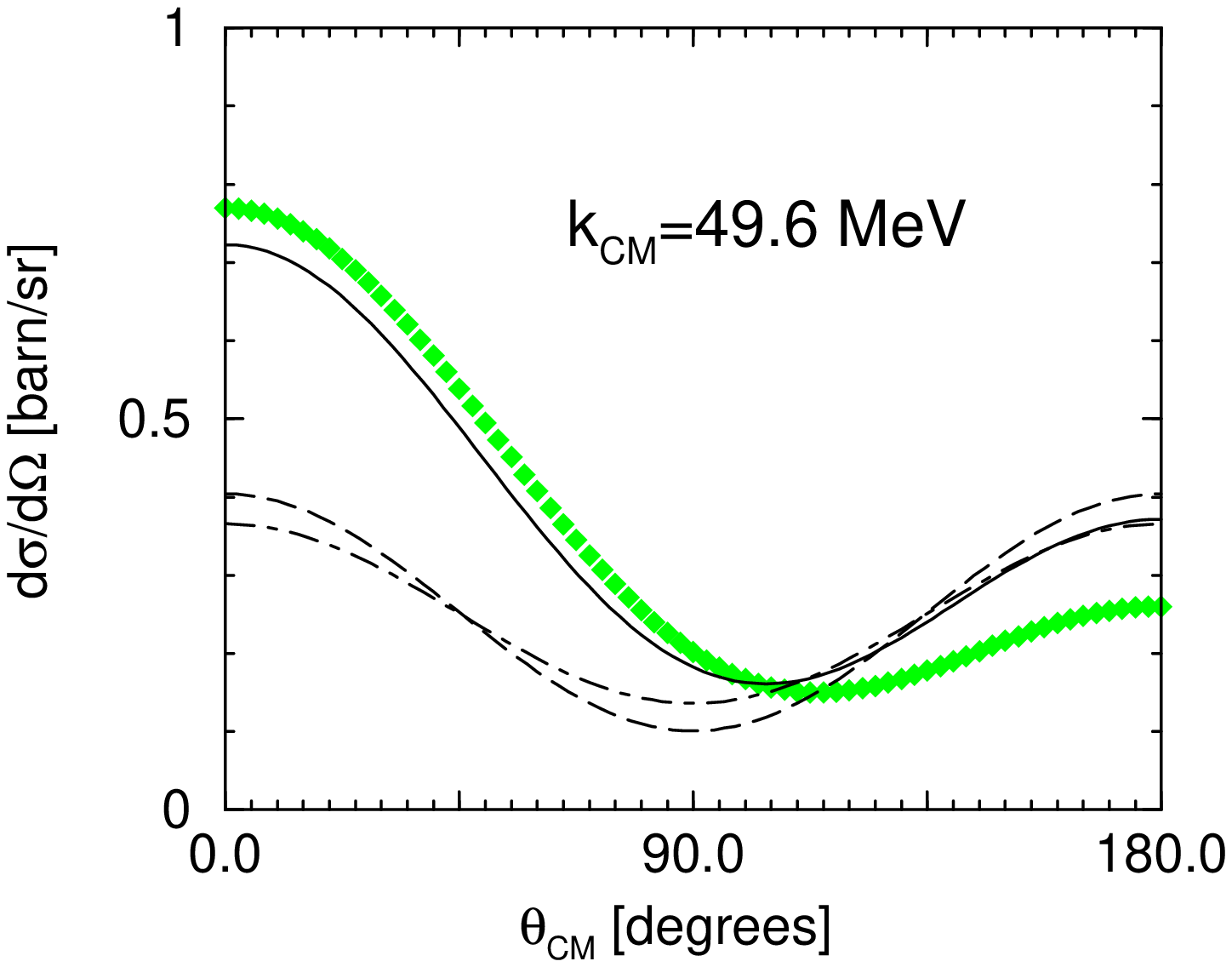,width=8cm,clip=true}}
\end{minipage}
\caption{Cross sections for $n\alpha$ scattering around the $p_{3/2}$
 resonance.
 Left: total cross section (in barns) as a function of the neutron kinetic 
 energy  (in MeV) in the $\alpha$ rest frame. 
 Right: differential cross section (in barns/sr) 
 as a function of the center-of-mass scattering angle $\theta_{CM}$ 
 (in degrees) at a center-of-mass momentum
 of $k_{CM}=49.6$ MeV. 
 The diamonds are evaluated data \protect\cite{BNL}, 
 and the black squares are experimental 
 data \protect\cite{data}.
 The dashed and solid lines show the EFT resummed results at 
 LO and NLO, respectively. (The dash-dotted line shows the  LO
 result in a modified power counting where the $1-$ partial 
 wave is promoted to leading order.)
From Ref.~\protect\cite{halos}.}
\label{fig:sigs2}
\end{figure}

The reason for this failure is easy to understand.
In a momentum region of ${\cal O}(\Mlo^{2}/\Mhi)$ around the resonance
there is a cancellation in the denominator of the $1+$ propagator,
bubbles have to be resummed to all orders, 
and the $1+$ propagator is enhanced by a factor of ${\cal O}(\Mhi/\Mlo)$.
In this region,
the $N\alpha$ $T$ matrix still has
the form (\ref{T}) but now
$\nu= \sum_i V_i d_i -3p +L$.
The corresponding 
results for the total and differential cross sections
with data in Fig. \ref{fig:sigs2}.
The success of this resummed NLO description is evident
throughout the low-energy region.
(Note that an additional resummation of the $1-$ propagator, which
would be necessary if there was a shallow resonance
in this channel, as sometimes claimed, does not seem
to improve the results significantly.) 
The description of the phase shifts themselves also comes out pretty well,
see Ref. \cite{halos}.

\section{OUTLOOK}

I have considered here only nuclear shallow states.
However, the ideas discussed above are much more general.
These EFTs can 
immediately be extended to other physical systems that contain
shallow bound states, such as certain molecules \cite{3bosons,4bosons,atoms}.
Apart from specific applications, these EFTs have two 
interesting generic features that lend them some intrinsic
mathematical interest as well.
First, they are the simplest theories where short-distance physics
produces non-perturbative structures at low energy.
Second, the non-perturbative character of the resulting renormalization
shows unique features, such as limit cycles.
Much more work is possible along these lines 
---see, {\it e.g.} Ref. \cite{limit}.

More within the scope of this conference,
there is certainly reason to push these EFTs in the direction of 
heavier nuclei.
This is in fact just the very beginning of halo EFT.
The next step is to use the $N\alpha$ interactions determined
from $N\alpha$ scattering \cite{halos} and the $NN$ interactions
determined
from $NN$ scattering \cite{ARNPSreview,gautam} to calculate the 
halo $^6$He as a $^4 {\rm He}+n+n$ system \cite{hans}, 
the same way we successfully described 
triton as a $p+n+n$ system \cite{pionlessT}.
But clearly the theory can be applied to reactions
involving any halo nucleus.
For example, to the extent that $^8$B can be regarded as a halo,
the reaction $p + \, ^7{\rm Be} \to \, ^8{\rm B} +\gamma$
can be analyzed as was $p+n\to d+\gamma$ \cite{ARNPSreview,gautam}.
This could become, I hope, a useful, systematic approach to physics near the
driplines.

\section*{Acknowledgments}

I am grateful to Paulo Bedaque, Carlos Bertulani and especially Hans Hammer
for enjoyable collaborations on the research reported here,
and to Bj\"orn Jonson, Bo H\"oistad and Dan Riska for the invitation to such a 
wide-ranging conference.

\end{document}